\documentclass[a4paper]{jpconf}
\usepackage{graphicx}
\usepackage{amsmath,amstext,amsthm,amssymb}
\usepackage{isotope}
\usepackage{bm}

\newcommand{\ve}[1]{\ensuremath{\mathbf{#1}}}

\newcommand{\Ep}{\ensuremath{E_N}}

\def\lsim{\lesssim}
\def\gsim{\gtrsim}

\begin{document}


\title{$\gamma$-ray production in neutral-current neutrino-oxygen
interactions at energies above 200 MeV
}
\author{A M Ankowski$^{1, 2}$, O Benhar$^1$, T Mori$^3$, R Yamaguchi$^3$ and M Sakuda$^3$}

\address{$^1$ INFN and Department of Physics,``Sapienza'' Universit\`a di Roma, I-00185 Roma, Italy}
\address{$^2$ On leave from Institute of Theoretical Physics, University of Wroc{\l}aw, Wroc{\l}aw, Poland}
\address{$^3$ Department of Physics, Okayama University, Okayama 700-8530, Japan}

\ead{Artur.Ankowski@roma1.infn.it}

\date{\today}%

\begin{abstract}
We report the results of a calculation of the neutrino- and antineutrino-induced $\gamma$-ray production cross section
for oxygen target. Our analysis is focused on the kinematical region of neutrino energy larger than $\sim$200MeV, in which single-nucleon
knockout is known to be the dominant reaction mechanism. The numerical results have been obtained using
a realistic model of the target spectral function, extensively tested against electron-nucleus scattering data.
We find that at neutrino energy 600~MeV the fraction of neutral-current interactions leading to emission of $\gamma$-rays of energy larger than 6~MeV is $\sim$41\%, and that
the contribution of the $p_{3/2}$ state is overwhelming.
\\\\
{\noindent}{\it Invited paper to NUFACT 11, XIIIth International Workshop on Neutrino Factories, Super beams and Beta beams, 1-6 August 2011, CERN and University of Geneva\\
(Submitted to IOP conference series)}
\end{abstract}

%
%
%
%



The observation of $\gamma$ rays originating from nuclear deexcitation can be exploited to identify neutral-current (NC)
neutrino-nucleus interactions in a broad energy range.
Following the pioneering studies of nuclear excitations by neutral weak currents of Refs.~\cite{Donnelly,Donnelly&Peccei},
theoretical calculations of the cross section of $\gamma$-ray production from NC neutrino-oxygen interactions
have been carried out in the neutrino energy range $E_\nu\sim 10$--500 MeV~\cite{KLV,Kolbe1,Kolbe2}.

Neutrons, while providing  $\sim$50\% of NC events, do not emit Cherenkov light. As a~conse{\-}quence, the availability
of an alternative signal allowing one to identify NC interactions is very important. Events with $\gamma$ rays of energy above the observational threshold of  5~MeV can be detected in a water Cherenkov detector,
like Super-Kamiokande, and contribute up to $\sim$5\% of the total event number~\cite{Beacom,T2K}, independent of neutrino oscillations. Note that in water $\sim$90\% (16 out of 18) of the NC interactions take place in oxygen.

At low energy, elastic scattering and inelastic excitation of discrete nuclear states provide the main contribution to
the neutrino-nucleus cross section. However,  at $E_\nu \gsim 200$~MeV the cross section associated with these processes
tends to saturate, and quasielastic (QE) nucleon knockout becomes the dominant reaction mechanism. If the residual nucleus is
left in an excited state, these processes can also lead to $\gamma$-ray emission.

\begin{figure}
\begin{minipage}[l]{0.46\textwidth}
\vspace{1.5em}
\centering
\includegraphics[width=0.72\textwidth]{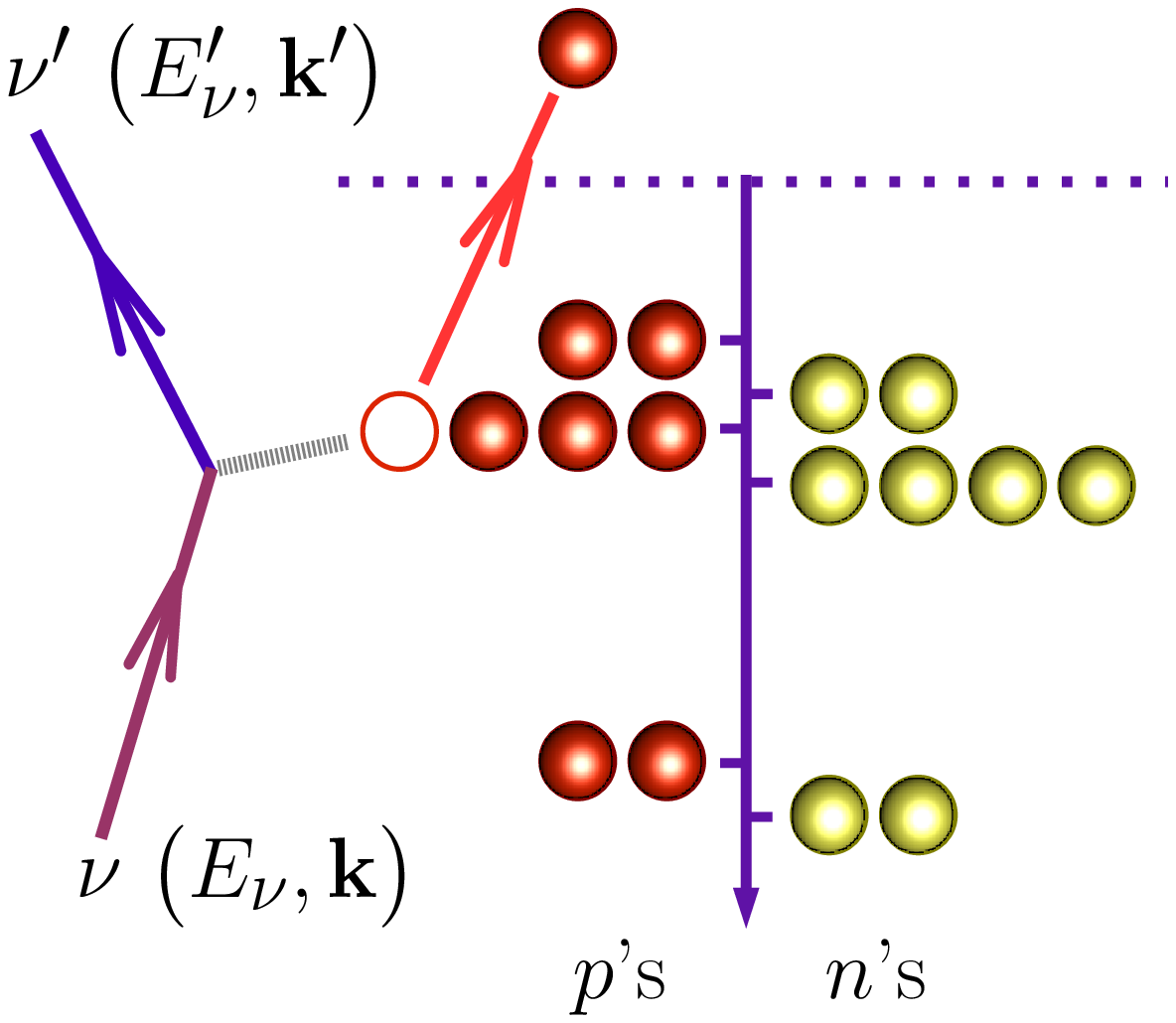}
\caption{\label{fig1} Schematic representation of neutral-current neutrino scattering off \isotope[16][8]{O}.}
\end{minipage}
\begin{minipage}[c]{0.08\textwidth}
\phantom{00000}
\end{minipage}
\begin{minipage}[r]{0.46\textwidth}
\centering
\includegraphics[width=0.96\textwidth]{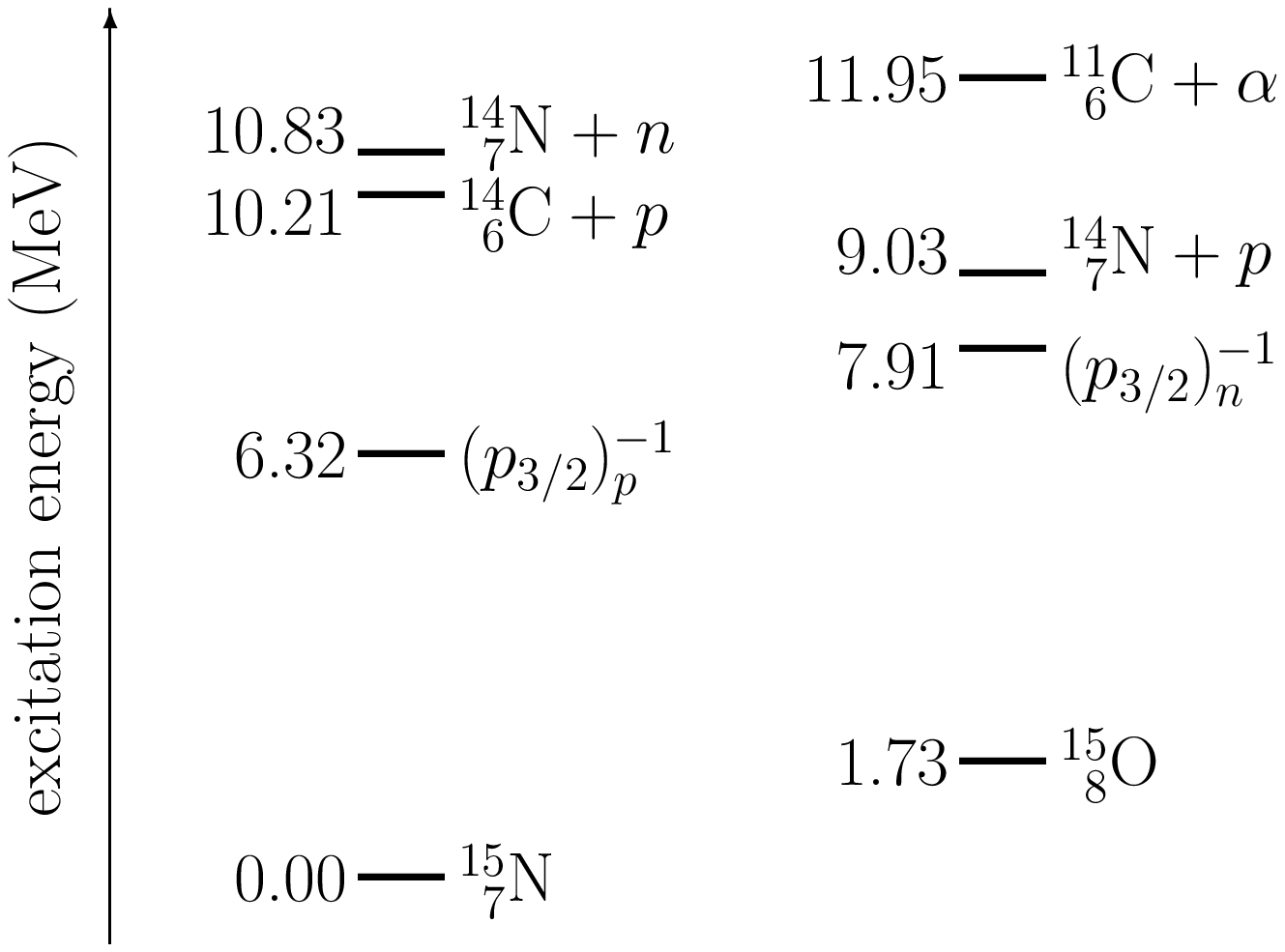}
 \caption{\label{fig2} Low-lying excited levels of the residual nuclei produced in $\isotope[16][8]{O}(\nu,\nu'N)$ scattering.} 
\end{minipage}
\end{figure}

In the QE regime, neutrino-nucleus scattering reduces to the incoherent sum of
elementary scattering processes involving individual nucleons, the energy and momentum of which are distributed according to the
target spectral function~\cite{RMP}. A schematic representation of NC QE neutrino-nucleus scattering is given in Fig.~\ref{fig1}, where the dashed line
represents the threshold for nucleon emission in the continuum.

In this paper, we discuss the emission of $\gamma$ rays
arising from the decay of the residual nuclei of the reactions
$\nu+\isotope[16][8]{O}\rightarrow \nu+p+\isotope[15][7]{N}^*$, or $\nu+\isotope[16][8]{O}\rightarrow \nu+n+\isotope[15][8]{O}^*$,
the cross sections of which have been computed using a realistic model of the oxygen spectral function.
Note that the  presented formalism applies also to antineutrino-induced $\gamma$-ray production.

In our approach, covered in detail in Ref.~\cite{Gamma}, the cross section of $\gamma$-ray production following a NC QE interaction, $\sigma_\gamma$,  is written in the form
\begin{equation}\label{eq:gammaCS}
\sigma_\gamma\equiv\sigma(\nu + \isotope[16][8]{O} \to \nu + \gamma +Y +N) = \sum_\alpha\sigma(\nu + \isotope[16][8]{O} \to \nu+X_\alpha + N )\:\textrm{Br}(X_\alpha \to \gamma +Y),
\end{equation}
where $N$ is the knocked out nucleon, $X_\alpha$ denotes the residual nucleus in the state $\alpha$,
and $Y$ is the system resulting from the electromagnetic decay of $X_\alpha$, e.g. \isotope[15][8]{O}, \isotope[15][7]{N},
$\isotope[14][7]{N} +n$, or $\isotope[14][6]{C} +p$~\cite{Isotopes1, Isotopes2}. The energy spectrum of the states of the
residual nuclei is schematically illustrated in Fig.~\ref{fig2}. The branching ratios
Br$(X_\alpha \to \gamma +Y)$  have been taken from Ref.~\cite{Ejiri&Kamyshkov}.

The NC QE cross section, $\sigma(\nu + \isotope[16][8]{O} \to \nu+X_\alpha + N)$, has been
calculated within the approach discussed in Ref.~\cite{Benhar1} for the case of charged-current interactions. It may be expressed as
\begin{eqnarray}
\frac{d\sigma_{\nu A}}{d\Omega dE'_\nu }= \sum_{N=p,\,n}\int d^3p\,
dEP_N({\bf p},E) \frac{M}{\Ep}\frac{d\sigma _{\nu N}}{d\Omega dE'_\nu },
\label{NC:xsec}
\end{eqnarray}
where $\Ep = \sqrt{M^2 + \ve p^2}$, $M$ being the nucleon mass,
$d\sigma _{\nu N}/d\Omega dE^\prime_\nu $ denotes the elementary neutrino-nucleon cross section and
the spectral function $P_N({\bf p},E)$ yields the probability of removing a nucleon of momentum ${\bf p}$ from the
target leaving the residual nucleus with energy $E + E_0 -M$, $E_0$ being the target ground-state energy.

According to the shell model, nuclear dynamics can be described by a mean field.
In the simplest implementation of this model, protons in the $\isotope[16][8]{O}$ nucleus occupy three single-particle states, $1p_{1/2}$, $1p_{3/2}$, and $1s_{1/2}$, with
removal energy 12.1, 18.4, and $\sim$42~MeV, respectively~\cite{Saclay,Nikhef,JLab}. The neutron levels exhibit the same pattern, see Fig.~\ref{fig1}, but are more deeply bound by 3.54~MeV~\cite{Isotopes2}.

As a consequence of a mean field dynamics, knock out of a target nucleon leaves the residual system in a bound state, and the spectral function can be
conveniently written in the form
\begin{equation}
P_N({\bf p},E) = \sum_{\alpha\: \in \{F\} } n_\alpha |\phi_\alpha({\bf p})|^2 f_\alpha (E-E_\alpha),
\label{S:MF}
\end{equation}
where $\phi_{\alpha}({\bf p})$ is the momentum-space wave function associated with the $\alpha$-th
shell-model state and the sum is extended to all occupied states belonging to the Fermi sea $\{F\}$.
The {\it occupation probability} $n_\alpha \leq 1$ and the (unit-normalized) function $f_\alpha(E-E_\alpha)$, describing
the energy width of the $\alpha$-th state, account for the effects of nucleon-nucleon (NN) correlations,
not included in the mean field picture.
In the absence of correlations, $n_\alpha \rightarrow 1$ and $f_\alpha(E-E_\alpha) \rightarrow \delta(E-E_\alpha)$.


A realistic model of the proton spectral function of oxygen has been obtained within the
local density approximation (LDA), combining the experimental data of
Ref.~\cite{Saclay} with the results of theoretical calculations of the correlation contribution
in uniform nuclear matter at different densities \cite{Benhar1,Benhar2}.
The results reported in Ref.~\cite{Benhar1} show that the LDA spectral function provides an
accurate description of the inclusive electron-oxygen cross sections at beam energies around
1~GeV. In addition, it predicts a nucleon momentum distribution in agreement with that obtained from
the data of Ref.~\cite{Daniela}.

As pointed out in Ref.~\cite{Benhar_90}, nucleon-knockout experiments measure {\em spectroscopic strengths}, not occupation probabilities.
Spectroscopic strengths are given by the area below the sharp peaks observed in the missing-energy spectra, corresponding to knockout of a nucleon occupying one of the
shell-model states, corrected to take into account final-state interactions. On the other hand, occupation probabilities include contributions corresponding to
larger removal energy, arising from mixing of the one-hole state with more complex final states~\cite{Benhar_90}.



\begin{table}[b]
\centering
    \begin{tabular}{@{}l|lll@{}}
    \br
    $\alpha$  &  $p_{1/2}$  &  $p_{3/2}$ & $s_{1/2}$\\[3pt]
    \hline
    $S_\alpha$  & 0.632 & \phantom{00}0.703 & \phantom{0}0.422\\
    Br$(X_\alpha\to \gamma+Y)$  & 0\% & 100\% & $16\pm1$\%\\
    \br
    \end{tabular}
\caption{\label{tab1} Spectroscopic strengths of the \isotope[16][8]{O} hole states and their branching ratios for deexcitation by the $E_\gamma>6$ MeV photon emission.}
\end{table}

\begin{figure}[b]
\centering
\includegraphics[width=0.5\columnwidth]{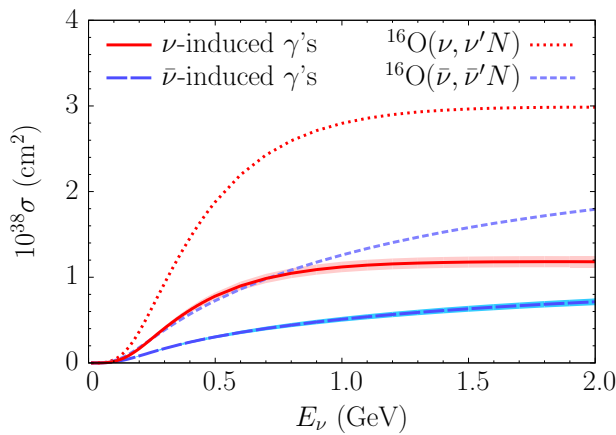}
\caption{\label{fig4} Cross section for $\gamma$-ray production following NC QE interaction of neutrino (solid line) and antineutrino (long-dashed line) compared to the NC QE cross section of neutrino (dotted line) and antineutrino (short-dashed line). Only the photons of energy larger that 6~MeV are considered.
}
\end{figure}

The $p_{1/2}$,  $p_{3/2}$ and $s_{1/2}$ spectroscopic strengths have been computed by integrating
the oxygen spectral function of Refs.~\cite{Benhar1,Benhar2} over the energy ranges
$11.0 \leq E \leq 14.0$ MeV, $17.25 \leq E \leq 22.75$ MeV, and $22.75 \leq E \leq 62.25$ MeV, respectively.
Dividing these numbers by the degeneracy of the shell-model states, one obtains the quantities $S_\alpha$  listed in Table~\ref{tab1}.
The same spectroscopic strengths have been used for protons and neutrons.


The branching ratios $\textrm{Br}(X_\alpha \to \gamma +Y)$, necessary to calculate the cross section $\sigma_\gamma$ according to
Eq.~\eqref{eq:gammaCS}, are collected in Table~\ref{tab1}~\cite{Ejiri&Kamyshkov}. In the case of the $p_{1/2}$-proton (neutron) knockout, the residual nucleus is \isotope[15][7]{N} (\isotope[15][8]{O}) produced in its ground state. Hence, no $\gamma$-rays are produced. As the $p_{3/2}$-proton (neutron) hole lies below the nucleon-emission threshold,  10.21~MeV (7.30~MeV), it always deexcites through photon emission with half-life $0.146\pm0.008$ fs (less than 1.74 fs)~\cite{Isotopes2}. When a proton (neutron) is knocked out from the deepest $s_{1/2}$ shell, the excitation energy is high enough for many deexcitation channels to open, of which only two, $\isotope[14][6]{C}+p$ and $\isotope[14][7]{N}+n$ ($\isotope[14][6]{C}+p$ and $\isotope[11][6]{C}+\alpha$), yield photons of energy higher than 6~MeV~\cite{Ejiri&Kamyshkov} (see Fig.~\ref{fig2}).


In Fig.~\ref{fig4} our results for the neutrino- and antineutrino-induced $\gamma$-ray production cross section are compared to the
neutrino and antineutrino NC QE cross sections. The error bands show the uncertainties arising form the determination of the spectroscopic strengths (5.4\%),
the treatment of Pauli blocking (1\%), and the branching ratio of the $s_{1/2}$ state (1\%).
The $\sigma_\gamma$'s dependence on neutrino energy is very similar, although not identical, to that of the NC QE cross section. The discrepancy arises from
 difference between the average removal energy associated with the whole spectral function and the energy of the $p_{3/2}$ shell, yielding the overwhelming
 contribution to $\sigma_\gamma$. The neutrino-induced $\gamma$-production cross section reaches its maximum at $E_\nu\sim1.9$ GeV and is slowly decreasing at larger energies. On the other hand, the corresponding antineutrino cross section is an increasing function of $E_\nu$.

In conclusion, we have computed the neutrino- and antineutrino-induced $\gamma$-ray production cross section
for oxygen target, focussing on the kinematical region in which single-nucleon knockout dominates. Considering photons of energy larger than 6~MeV, we find that
the $p_{3/2}$ state provides the overwhelming contribution, and that the ratio $\sigma_\gamma/\sigma_{{\rm NC}}$,
exhibiting a significant energy-dependence at $E_\nu \lsim 1$ GeV, is $\sim$41\% at $E_\nu=600~$MeV. Our results,
obtained using a realistic model of the target spectral function, provide an accurate estimate of a signal that
can be exploited to identify neutral-current events in water-Cherenkov detectors.

\ack
This work was supported by INFN (Grant MB31), MIUR PRIN (Grant ``Many-body theory of nuclear systems and implications on the
physics of neutron stars''), and in part by the Grant-In-Aid from the Japan Society for Promotion of Science (Nos. 21224004 and 23340073).
A.M.A. was supported by the Polish Ministry of Science and Higher Education under Grant No. 550/MOB/2009/0.

\end{document}